\documentclass[fleqn,usenatbib,useAMS]{mnras}

\usepackage{graphicx}	% Including figure files
\usepackage{amsmath}	% Advanced maths commands
\usepackage{amssymb}	% Extra maths symbols
\usepackage{multicol}        % Multi-column entries in tables
\usepackage{bm}		% Bold maths symbols, including upright Greek
\usepackage{pdflscape}	% Landscape pages

\newcommand{\cmlt}{$\alpha_{\rm MLT}$} 
\newcommand{\acen}{$\alpha$ Cen}
\newcommand{\sad}{$s_{\rm ad}$}

\defcitealias{Spada_ea:2018}{Paper I}

\def\apj{{ApJ}}                 % Astrophysical Journal
\def\apjl{{ApJ}}                % Astrophysical Journal, Letters
\def\apjs{{ApJS}}               % Astrophysical Journal, Supplement
\def\aap{{A\&A}}                % Astronomy and Astrophysics
\def\apss{{Ap\&SS}}          % Astrophysics and Space Science
\def\mnras{{MNRAS}}             % Monthly Notices of the RAS
\def\ssr{{Space~Sci.~Rev.}}
\usepackage[T1]{fontenc}
\usepackage{ae,aecompl}

\usepackage{newtxtext,newtxmath}
\title[Entropy-calibrated models of $\alpha$ Cen A and B]{Testing the entropy calibration of the radii of cool stars: \\ models of $\alpha$ Centauri A and B}
\author[F. Spada \& P. Demarque]{Federico Spada$^{1}$\thanks{Contact e-mail: \href{mailto:spada@mps.mpg.de}{spada@mps.mpg.de}} and Pierre Demarque$^{2}$ \\
$^{1}$Max-Planck Institut f\"ur Sonnensystemforschung, Justus-von-Liebig Weg 3, 37077 G\"ottingen, Germany\\
$^{2}$Department of Astronomy, Yale University, New Haven, CT 06520-8101, USA}

\date{Last updated ; in original form}
\pubyear{}

\begin{document}
\label{firstpage}
\pagerange{\pageref{firstpage}--\pageref{lastpage}}
\maketitle
%
%%%%%%%%%%%%%%%%%%%%%%%%%%%%%%%%%%%%%%%%%%%
\begin{abstract}
% Note: must be < 250 words long
We present models of $\alpha$ Centauri A and B implementing an entropy calibration of the mixing-length parameter \cmlt{}, recently developed and successfully applied to the Sun (Spada et al. 2018, ApJ, 869, 135).
In this technique the value of \cmlt{} in the 1D stellar evolution code is calibrated to match the adiabatic specific entropy derived from 3D radiation-hydrodynamics simulations of stellar convective envelopes, whose effective temperature, surface gravity, and metallicity are selected consistently along the evolutionary track. 

The customary treatment of convection in stellar evolution models relies on a constant, solar-calibrated \cmlt{}. 
There is, however, mounting evidence that this procedure does not reproduce the observed radii of cool stars satisfactorily. 
For instance, modelling \acen{} A and B requires an ad-hoc tuning of \cmlt{} to distinct, non-solar values.

The entropy-calibrated models of \acen{} A and B reproduce their observed radii within $1\%$ (or better) without externally adjusted parameters.
The fit is of comparable quality to that of models with freely adjusted \cmlt{} for \acen{} B (within $1$ $\sigma$), while it is less satisfactory for \acen{} A (within $\approx 2.5$ $\sigma$).  
This level of accuracy is consistent with the intrinsic uncertainties of the method.

Our results demonstrate the capability of the entropy calibration method to produce stellar models with radii accurate within $1\%$. 
This is especially relevant in characterising exoplanet-host stars and their planetary systems accurately.
\end{abstract}

%%%%%%%%%%%%%%%%%%%%%%%%%%%%%%%%%%%%%%%%%%%

%%%%%%%%%%%%%%%%%%%%%%%%%%%%%%%%%%%%%%%%%%%%
\section{Introduction}
\label{introduction}
%%%%%%%%%%%%%%%%%%%%%%%%%%%%%%%%%%%%%%%%%%%%
%- observational evidence for non-solar \cmlt{}
%- brief review of alternative \cmlt{} calibration approaches
%- summary of paper I
%- short summary of papers about \acen{} modeling (classical vs. asteroseismic constraints)

In the outer layers of solar-like, or ``cool'' stars (mass $M_* \lesssim 1.3 \, M_\odot$), convective energy transfer is triggered because of the large radiative opacity.
The presence of an outer convection zone affects significantly the overall structure of the star \citep[e.g.,][]{Schwarzschild:1958,Clayton:1968,KWW12}, and sustains the generation of magnetic fields through dynamo action and thus magnetic activity \citep[see][for a review emphasising the solar-stellar connection]{Engvold_ea:2019}.

The treatment of convection is a long-standing issue in modelling cool stars.
In particular, no satisfactory theoretical prescription exists to link the deeper portion of the convective envelope, where efficient convection maintains an essentially adiabatic temperature stratification, with the outer sub-photospheric layers, which are dominated by radiative losses and characterised by super-adiabatic stratification. 
The non-trivial structure of this super-adiabatic layer (SAL) determines the value of the entropy in the deep, adiabatic convection zone, $s_{\rm ad}$.
The quantity $s_{\rm ad}$, in turn, directly controls the overall depth of the convection zone and the radius of the star \citep[e.g.,][]{Schwarzschild:1958,HansenKawalerTrimble:2004}. 
 
The mixing-length theory (MLT; \citealt{BV58}) is almost universally adopted to model the transition from the SAL to the adiabatic interior in standard stellar evolution codes, even if some of its basic assumptions are known from detailed numerical simulations to be not realistic \citep[see][for details]{Trampedach:2010}. 
In addition, the MLT contains a free parameter, \cmlt{}, that requires external calibration.
This is usually accomplished exploiting the uniquely favourable case of the Sun, for which simultaneous independent knowledge of the mass, radius, and age is available \citep[``standard solar model calibration"; see, e.g.,][]{Basu_Antia:2008}.
The solar-calibrated value of \cmlt{} is thus used to model stars of any mass and chemical composition, and kept constant throughout their entire evolution.
The assumption of universality of \cmlt{} is, however, neither theoretically justified, nor supported by observations; indeed, there is mounting evidence to the contrary (e.g., \citealt{Lebreton_ea:2001, Yildiz_ea:2006, Bonaca_ea:2012, Tayar_ea:2017, Joyce_Chaboyer:2018a, Viani_ea:2018}; but see also \citealt{Valle_ea:2019}).

For these reasons, since when Radiation Hydro-Dynamics (RHD) numerical simulations of convection have become available, several attempts have been made to incorporate their results in 1D stellar models to improve upon the MLT treatment.
Early works focused on using the RHD simulations to calibrate the \cmlt{} parameter as a function of its metallicity and position in the $(\log\, T_{\rm eff}, \log\, g)$  Kiel diagram \citep{Ludwig_ea:1995,Ludwig_ea:1998,Ludwig_ea:1999}.
More recently, \citet{Trampedach_ea:2014a, Trampedach_ea:2014b} have extracted from their 3D RHD simulations horizontal and temporal averages of the $T$--$\tau$ relation, and an effective \cmlt{} variation.
These authors have also emphasised the importance of ensuring consistency with the microphysics of the RHD simulations when implementing these prescriptions in a 1D stellar evolution code.
Stellar evolution tracks for cool stars implementing the opacities, $T$--$\tau$ relations, and \cmlt{} calibration of \citet{Trampedach_ea:2014a, Trampedach_ea:2014b} have been constructed for the first time by \citet{Salaris_Cassisi:2015} and, more recently, by \citet{Mosumgaard_ea:2017, Mosumgaard_ea:2018}.
Even stricter adherence to the 3D simulations, at the price of increased complexity of implementation, can be achieved by replacing the envelope calculated from the 1D stellar code with an averaged profile extracted from 3D simulations (``patching''). 
Both static and evolutionary (``on the fly") patching implementations have been
developed \citep[see][respectively]{Jorgensen_ea:2017, Jorgensen_ea:2018}.

A novel approach, recently formulated by \citet{Tanner_ea:2016}, shifts the emphasis from \cmlt{} to \sad{} as the fundamental quantity used to link the RHD simulations with the 1D stellar models.
These authors have noted that the adiabatic specific entropy predicted by the RHD simulations is remarkably insensitive to the choices of the input physics.
The value of \sad{}, on the other hand, uniquely determines the radius of the 1D stellar model \citep{HansenKawalerTrimble:2004}, and can be readily calculated using the standard MLT equations.
To construct entropy-calibrated evolutionary tracks, the MLT parameter is calibrated at each step to reproduce in the 1D model the adiabatic specific entropy obtained from the RHD simulations as a function of the current metallicity and position in the Kiel diagram.
\citeauthor{Spada_ea:2018} (\citeyear{Spada_ea:2018}; hereafter \citetalias{Spada_ea:2018}) have implemented this procedure in the Yale stellar evolution code, and successfully applied it to construct a realistic solar model without the need to adjust the parameter \cmlt{}.

The entropy calibration has the advantage of retaining the simplicity of implementation of the MLT formalism, without requiring strict consistency with the microphysics of the RHD simulations.
Although this approach is not suitable when accurate modelling of the stratification of the outer layers is needed, it is adequate to establish a more realistic calibration of the radii of cool stars.

In  \citetalias{Spada_ea:2018} we showed that the entropy calibration has a significant impact on the stellar radius and the depth of the convection zone, and, to a lesser extent, on the chemical composition profile of the interior, as a result of the modified convective mixing history.
On the contrary, the central layers, the nuclear energy generation, and the luminosity are essentially unaffected.
The largest changes in the evolutionary track occur in the vicinity of the Hayashi line, i.e., during the early pre-main sequence and during the red giant branch phases.

In this paper we present a new application of the entropy calibration method.
After the successful test of constructing a realistic solar model, a natural follow-up case study is the $\alpha$ Centauri system.
The two more massive components of the system have masses bracketing the mass of the Sun and a moderately but significantly non-solar composition.
Moreover, their observational parameters are known with exquisite precision.
The literature on modelling \acen{} is vast; recent comprehensive reviews can be found in the introductions of \citet{Eggenberger_ea:2004} and of \citet{Joyce_Chaboyer:2018b}.

Most recently, \citet{Joyce_Chaboyer:2018b} have constructed a large grid of models of \acen{} A and B, taking into account a range of assumptions on the modelling parameters and input physics.
Their best-fitting models require the parameter \cmlt{} to be adjusted to a non-solar value, different for the two stars, in agreement with the results of previous analyses.
A specific aim of the present paper is therefore to test whether such difference in \cmlt{} for \acen{} A and B can be predicted by our entropy calibration.

This paper is organized as follows: we describe our standard and entropy-calibrated models of \acen{} A and B in Section \ref{methods}; our results are presented in Section \ref{results}, and discussed in Section \ref{discussion}; we summarise our conclusions in Section \ref{conclusions}.

%%%%%%%%%%%%%%%%%%%%%%%%%%%%%%%%%%%%%%%%%%%%
\section{Methods}
\label{methods}
%%%%%%%%%%%%%%%%%%%%%%%%%%%%%%%%%%%%%%%%%%%%

\subsection{The stellar evolution code}

All the models were constructed using the YREC stellar evolution code in its non-rotational configuration \citep{Demarque_ea:2008}. 
The details of the input physics are summarised as follows \citepalias[cf.][]{Spada_ea:2018}.
We use the OPAL 2005 equation of state \citep{Rogers_Nayfonov:2002}, and the OPAL opacities \citep{Rogers_Iglesias:1995,Iglesias_Rogers:1996}, supplemented by the low-temperature opacities of \citet{Ferguson_ea:2005} at $\log\, T \leq 4.5$. 
Diffusion of helium and heavy elements is taken into account according to the formulation of \citet{Thoul_ea:1994}. 
In the atmosphere, the Eddington grey $T$--$\tau$ relation is used. 
Abundances of elements are scaled to the \citet{Grevesse_Sauval:1998} solar mixture, with $(Z/X)_\odot=0.0230$.

The standard treatment of convection in YREC is based on the MLT \citep{BV58}, with the value of \cmlt{} calibrated on the Sun or freely adjusted, and assumed constant along the evolutionary track.
In the following, we refer to constant-\cmlt{} models as ``standard".
In \citetalias{Spada_ea:2018} we introduced a prescription to calibrate \cmlt{} using the adiabatic specific entropy derived from 3D RHD simulations of convection, and to update its value consistently with $\log\, g$ and $T_{\rm eff}$ at each evolutionary step.
We will refer to these models as ``entropy-calibrated".

\begin{table}
\caption{Adopted observational parameters of \acen{} A and B.}
\label{alpha_cen}
\begin{tabular}{lccc}
\hline
\hline
Parameter & \acen{} A & \acen{} B & References \\
\hline
Mass ($M_\odot$) & $1.1055 \pm 0.004$ & $0.9373 \pm 0.003$ & (1) \\
Radius ($R_\odot$) & $1.2234 \pm 0.0053$ & $0.8632 \pm 0.004$ & (1) \\
Luminosity ($L_\odot$) & $1.521 \pm 0.015$ & $0.503 \pm 0.006$ & (1) \\
Metallicity $Z/X$ & $0.039 \pm 0.006$ & $0.039 \pm 0.006$ & (2,3) \\  
\hline
\end{tabular}
(1): \citet{Kervella_ea:2017}; (2): \citet{Thoul_ea:2003};  (3): \citet{PortodeMello_ea:2008}.
\end{table}

\subsection{Standard models of \acen{} A and B}

We have constructed standard models of \acen{} A and B, to serve as an internally consistent term of comparison for our entropy-calibrated models.
We set up an optimisation process to find the values of the input parameters that produce the best fit of the observational constraints available for \acen{} A and B.
Our best-fitting procedure takes into account the ``classical'' parameters: mass, radius, luminosity, and surface metallicity of the two stars.
The asteroseismic parameters \citep{deMeulenaer_ea:2010,Kjeldsen_ea:2005} were not taken into account in the fit.  
The values of the observational parameters adopted for our fit with standard models are listed in Table \ref{alpha_cen}. 

It should be emphasised that constructing standard models of \acen{} A and B that are compatible with all the observational constraints currently available (i.e., classical and asteroseismic) is beyond the scope of the present work.
Rather, the purpose of our best fit with standard models is two-fold: 1) determine the values of \cmlt{} which reproduce the observed radii of the two stars, and in particular assess the significance of the requirement of non-solar, distinct \cmlt{} for \acen{} A and B previously found in the literature; 2) constrain their age and composition parameters (initial helium fraction $Y$ and bulk metallicity $Z$), which should be approximately the same for the two stars, since they are members of the same system.

The luminosity and surface metallicity at a given age are mostly determined by $Y$ and $Z$, respectively, and are very weakly sensitive to the entropy calibration of \cmlt{} \citepalias[see][]{Spada_ea:2018}. 
The constraints derived from the fit with standard models are therefore relevant to the entropy-calibrated models as well.
In this sense, comparing the fit of \acen{} A and B obtained with standard vs. entropy-calibrated models is a test of the entropy calibration almost as direct as in the solar case, which was presented in \citetalias{Spada_ea:2018}.   

\begin{figure}
\begin{center}
\includegraphics[width=0.5\textwidth]{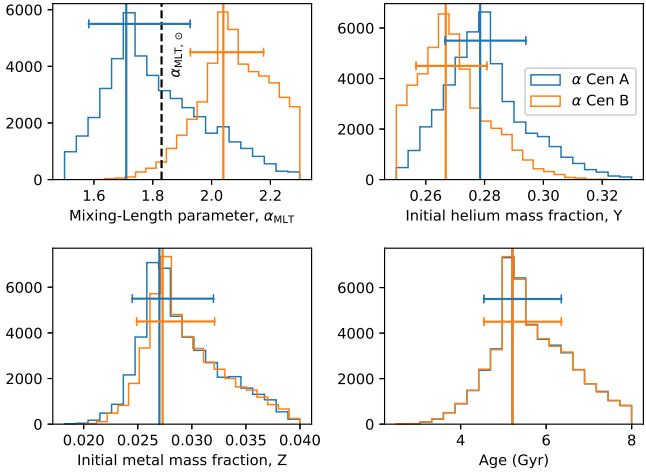}
\caption{Posterior probability distributions for the best-fitting parameters of the standard models of \acen{} A and B. Note that the fit requires distinct values of \cmlt{} for the two stars, and that $\alpha_{\rm MLT,B}$ is significantly non-solar. 
A moderate difference in the helium mass fraction is also visible.}
\label{acen_parms_fit}
\end{center}
\end{figure}

In our best-fit with standard models, we seek the optimal values of the MLT parameter, initial helium fraction, initial metallicity, and age of both stars: $\{ \alpha_{\rm MLT,A}$, $Y_A$, $Z_A$, $t_A$; $\alpha_{\rm MLT,B}$, $Y_B$, $Z_B$, $t_B\}$. 
The masses are kept fixed at their central observed values.
The fit takes into account the radius, luminosity, and surface metallicity of both stars $\{R_A$, $L_A$, $(Z/X)_A$; $R_B$, $L_B$, $(Z/X)_B\}$, as well as their binarity, by introducing a penalty for solutions that have too large values of $|Y_A-Y_B|$, $|Z_A-Z_B|$, or $|t_A-t_B|$.
The quality of the fit is evaluated in terms of the following chi-square function \citep[cf.][]{Joyce_Chaboyer:2018b}:
\begin{equation}
\chi^2_{\rm tot} =  \chi^2_{A, B} + \chi^2_{\rm bin},
\end{equation}
where the first term measures the fit of the two components individually, while the second term accounts for the binarity constraint:
\begin{eqnarray}
\nonumber
\chi_{A,B}^2 &= 
\dfrac{(R_{\rm A, obs}-R_{\rm A, mod})^2}{\sigma_{R_A}^2}
+ \dfrac{(L_{\rm A, obs}-L_{\rm A, mod})^2}{\sigma_{L_A}^2}
\\
\label{eq_chiAB}
&+ \dfrac{(R_{\rm B, obs}-R_{\rm B, mod})^2}{\sigma_{R_B}^2}
+ \dfrac{(L_{\rm B, obs}-L_{\rm B, mod})^2}{\sigma_{L_B}^2}
\\
\nonumber
&+ \dfrac{[(Z/X)_{\rm obs}-(Z/X)_{\rm mod}]^2}{\sigma_{Z/X}^2};
\\
\label{eq_chibin}
\chi^2_{\rm bin} &= \dfrac{(t_A-t_B)^2}{\sigma_{\Delta t}^2} +
                               \dfrac{(Y_A-Y_B)^2}{\sigma_{\Delta Y}^2} +
                               \dfrac{(Z_A-Z_B)^2}{\sigma_{\Delta Z}^2}.
\end{eqnarray}
In the equations above, ``obs" and ``mod" stand for observed and modeled, respectively; $(Z/X)_{\rm mod} = \frac{1}{2} [(Z/X)_{\rm A, mod} + (Z/X)_{\rm B, mod}]$; $\sigma_{\Delta t}= 5$ Myr; $\sigma_{\Delta Y} = 5\cdot 10^{-3}$; $\sigma_{\Delta Z} = 5 \cdot 10^{-4}$.
With these definitions, the fit optimises eight parameters against eight independent constraints.

\begin{table}
\caption{Best-fitting parameters for the standard models of \acen{} A and B.}
\label{mcmc}
\begin{tabular}{lcccc|c}
\hline
\hline
Parameter & Mode & Median & $\Delta_+$ & $\Delta_-$ & Adopted \\
\hline
$\alpha_{\rm MLT, A}$  &  $1.710$    &  $1.779$     & $0.2171$ &  $0.1273$ & $1.71^{+0.22}_{-0.13}$ \\
$\alpha_{\rm MLT, B}$  &  $2.039$    &  $2.076$     & $0.1366$ &  $0.1120$  &  $2.04^{+0.14}_{-0.11}$ \\
$Y_A$                          &  $0.2785$  &  $0.2788$  & $0.0155$  & $0.0119$ & $0.2785^{+0.016}_{-0.012}$ \\
$Y_B$                          &  $0.2668$  &  $0.2677$  & $0.0140$  & $0.0101$ & $0.2668^{+0.014}_{-0.010}$ \\
$Z_A$                          &  $0.0270$  &  $0.0282$  & $0.0050$  & $0.0025$ &  $0.0270^{+0.0050}_{-0.0025}$ \\ 
$Z_B$                          &  $0.0273$  &  $0.0286$  & $0.0048$  & $0.0024$ &  $0.0273^{+0.0024}_{-0.0048}$ \\
$t_A$ (Gyr)                  &  $5.211$    &  $5.509$     & $1.1425$  & $0.6675$ & $5.21^{+0.67}_{-1.14}$ \\
$t_B$ (Gyr)                  &  $5.212$    &  $5.509$     & $1.1421$  & $0.6674$ & $5.21^{+0.67}_{-1.14}$\\
\hline
\end{tabular}
$\Delta_+=84^{\rm th}-50^{\rm th}$ percentile; $\Delta_-=50^{\rm th}-16^{\rm th}$ percentile.
\end{table}

Operationally, the fit was performed in two steps. 
A first estimate of the best-fitting values of the parameters was obtained using the \citet{Nelder_Mead:1965} simplex algorithm as implemented in the \texttt{optimize.minimize} function, which is part of the \texttt{SciPy}\footnote{\url{http://www.scipy.org/}} package.
Subsequently, we performed a Markov Chain Monte Carlo (MCMC) sampling of the posterior probability distribution to estimate the uncertainties on the best-fitting parameters.
We used the affine invariant MCMC sampler of \citet{Goodman_Weare:2010} implemented in the \texttt{emcee} package \citep{Foreman-Mackey_ea:2013}.
The MCMC sampling was run with $32$ walkers, each performing $1400$ steps (see Figure \ref{corner_plot}). 
Each step requires running YREC twice, with input parameters $\{ \alpha_{\rm MLT,A}$, $Y_A$, $Z_A$, $t_A\}$ for \acen{} A, and $\{\alpha_{\rm MLT,B}$, $Y_B$, $Z_B$, $t_B\}$ for \acen{} B. 

For each parameter, we adopt the mode of the posterior distribution as the best-fitting value.
The upper and lower uncertainties are estimated as the difference between the $84$-th and $50$-th percentiles, and the difference between the $50$-th and $16$-th percentiles, respectively.
The discrepancy between the mode and the median is a measure of the departure of the posterior probability distribution from a Gaussian distribution.
All the parameters exhibit significantly non-Gaussian distributions. 
Strong correlations between $Y_A$ and $Y_B$, $Z_A$ and $Z_B$, and $t_A$ and $t_B$ are also evident.
These correlations are introduced by the binary constraint (equation \ref{eq_chibin}). 

The probability distributions for the best-fitting parameters of our standard models of \acen{} A and B  are plotted in Figure \ref{acen_parms_fit}, and their values are reported in Table \ref{mcmc}.
From these results we can derive two main conclusions.
First, distinct values of $\alpha_{\rm MLT, A}$ and $\alpha_{\rm MLT, B}$ are required to achieve a satisfactory fit.
The best-fitting value of $\alpha_{\rm MLT, B}$ is also significantly non-solar (top left panel of Figure \ref{acen_parms_fit}).
Secondly, while the best-fitting metallicity and age are essentially identical for the two stars, the helium abundances are moderately discrepant, although not significantly so with respect to the error bars defined by the 84th and 16th percentiles of the distributions (top right panel).
Both conclusions are consistent with the results of \citet{Joyce_Chaboyer:2018b}.

\subsection{Entropy-calibrated models of \acen{} A and B}

The entropy-calibrated models were calculated using the same procedure discussed at length in \citetalias{Spada_ea:2018}, which is summarised as follows.
At each evolutionary time step, the MLT parameter in the 1D stellar model is adjusted so that the entropy at the bottom of the convection zone, \sad{}, matches its corresponding value obtained in the 3D RHD simulations of convection (see Figure 3 of \citetalias{Spada_ea:2018}).
This calibration is based on the functional dependence of \sad{} on the effective temperature, surface gravity, and metallicity of the star, which is expressed in analytic form as:
\begin{align}
\label{entropy}
s_{\rm ad}(x) &= s_0 + \, \beta \exp{\left(-\frac{x-x_0}{\tau}\right)} + s_{\rm offset};
\\
\label{xdef}
x&=A\, \log T_{\rm eff} + B\, \log g,
\end{align}
where the parameters $A$, $B$, $s_0$, $x_0$, $\beta$, $\tau$ depend on metallicity, and were determined by \citet{Tanner_ea:2016} via best-fitting of the simulations of \citet{Magic_ea:2013a,Magic_ea:2013b,Magic_ea:2015a,Magic_ea:2015b,Tanner_ea:2013a,Tanner_ea:2013b,Tanner_ea:2014}.
The additive constant $s_{\rm offset}$ is a small correction (of the order of $1\%$ of \sad{}), introduced in \citetalias{Spada_ea:2018} to achieve consistency with the standard solar model.

In this work, all the parameters in equations \eqref{entropy} and \eqref{xdef} are the same as listed in Table 1 of \citet{Tanner_ea:2016} and used in \citetalias{Spada_ea:2018}. 
It should be stressed that these parameters are entirely determined by the 3D RHD simulations. 
In particular, they were evaluated independently of the 1D stellar evolution code, except, of course, for the additive constant $s_{\rm offset}$. 
As was shown by \citet{Tanner_ea:2016}, the  $s_{\rm ad}(x)$ relations are very weakly sensitive to the input physics used in the 3D RHD simulations (e.g., \citealt{Tanner_ea:2013a} vs. \citealt{Magic_ea:2013a}). 
This property of $s_{\rm ad}$ is critical for the practical implementation of the entropy calibration method: it makes the determination of the parameters in equation \eqref{entropy} robust, and directly applicable to any 1D stellar evolution code.

The calibration of \cmlt{} is performed at each evolutionary time step, consistently with the current values of $\log \, g$, $T_{\rm eff}$, and surface metallicity.
This approach relies on the simplicity of the MLT framework, but suffices to specify more realistic surface boundary conditions and improve the radius calibration of the 1D stellar models.
The stratification of the outer layers, however, is still modelled using the MLT, and its accuracy is therefore not improved with respect to the standard approach.

We have constructed entropy-calibrated models of \acen{} A and B with the same composition and age as the standard models discussed in the previous subsection.
It should be emphasised that the evolution of \cmlt{} in the entropy-calibrated runs is entirely determined by the calibration of \cmlt{} against the adiabatic specific entropy.
In other words, the treatment of convection in the entropy-calibrated models contains no adjustable parameters.

\begin{table}
\centering
\caption{Standard vs. entropy-calibrated models of \acen{} A and B.}
\label{entr_vs_std}
\begin{tabular}{lccc}
\hline
\hline
Parameter & Observed & Standard & Entr.-cal. \\
\hline
& \multicolumn{3}{c}{\acen{} A} \\
\hline
$R/R_\odot$ & $1.2234\pm0.0053$ &  $1.2235$  &  $1.2097$   \\ 
$L/L_\odot$  & $1.521\pm0.015$     &  $1.5091$  &  $1.5045$   \\
$Z/X$            & $0.039\pm0.006$     & $0.0315$  &  $0.0348$   \\
%Mass ($M_\odot$) &  N/A                  & $1.055$    &  $1.055$    \\
Age   (Gyr)    &  N/A                          & $5.21$      &   $5.21$      \\
$\alpha_{\rm MLT}$  &  N/A              & $1.71$      &  $1.758$ \\
$s_{\rm ad}$  & N/A & $1.876\cdot 10^9$ &  $1.853\cdot 10^9$ \\
\hline
& \multicolumn{3}{c}{\acen{} B} \\
\hline
$R/R_\odot$ & $0.8632\pm0.004$ &  $0.8647$  &  $0.8612$    \\
$L/L_\odot$  & $0.503\pm0.007$   &  $0.5122$    &  $0.5107$    \\
$Z/X$            & $0.039\pm0.006$   &  $0.0346$  &  $0.0364$    \\
%Mass ($M_\odot$) &  N/A                  & $0.9373$  &  $0.9373$    \\
Age   (Gyr)    &  N/A                          & $5.21$      &   $5.21$      \\
$\alpha_{\rm MLT}$  &  N/A               & $2.04$      &  $2.045$ \\
$s_{\rm ad}$ & N/A & $1.649\cdot 10^9$ &  $1.639\cdot 10^9$ \\
\hline
\end{tabular}
\\
The units of adiabatic specific entropy are: erg g$^{-1}$K$^{-1}$.
\end{table}

%%%%%%%%%%%%%%%%%%%%%%%%%%%%%%%%%%%%%%%%%%%%
\section{Results}
\label{results}
%%%%%%%%%%%%%%%%%%%%%%%%%%%%%%%%%%%%%%%%%%%%

\subsection{\acen{} A and B as a test of the entropy calibration}

Entropy-calibrated and standard models of \acen{} A and B are compared in Table \ref{entr_vs_std}.
All the models are constructed using the best-fitting values of the composition and age parameters (and of \cmlt{}, for the standard models) listed in Table \ref{mcmc}.

As expected, the observable that is most affected by the entropy calibration of \cmlt{} is the radius.
For \acen{} A, the agreement of the entropy-calibrated model with the observed radius is within 2.5~$\sigma$, significantly worse than for the standard model with freely adjusted \cmlt{}, but still within $\approx 1\%$. 
For \acen{} B, the entropy-calibrated model reproduces the observed radius equally well as the standard model (within 1~$\sigma$, or $0.2\%$).

The impact of the entropy calibration on the luminosity is very modest, and as a result the fit of the observed luminosity is comparable to that of standard models for both stars.
The entropy-calibrated models reproduce the observed surface metallicity slightly better than the standard ones for both stars; this improvement is however not significant with respect to the relatively large observational uncertainties on $Z/X$.

Regarded as a test of the entropy calibration approach, the comparison with the standard models of \acen{} A and B is very encouraging. 
It shows that the entropy-calibrated models reproduce the observed radii of \acen{} A and B with an accuracy of $\approx 1\%$, or better, even if the freedom in adjusting the parameter \cmlt{} has been removed.

\subsection{Accuracy of the entropy calibration}

The performance of the entropy-calibrated models in reproducing the radius of \acen{} A and B can be put in a broader context, which illustrates the accuracy and limitations of the method.
Table \ref{entr_vs_std} reports \sad{} for the standard and entropy-calibrated models. 
For the latter, \sad{} is consistent with equation \eqref{entropy} by construction.
Note, however, that the adiabatic specific entropy of the standard models (with freely adjusted \cmlt{}) also agree within $\lesssim 1\%$ with equation \eqref{entropy}.
This is a consequence of the tight relation between \sad{} and the stellar radius, upon which the entropy calibration rests.
Since the radii of \acen{} A and B are known with high precision, this can also be interpreted as an independent test of equation \eqref{entropy}.

\begin{figure}
\begin{center}
\includegraphics[width=0.5\textwidth]{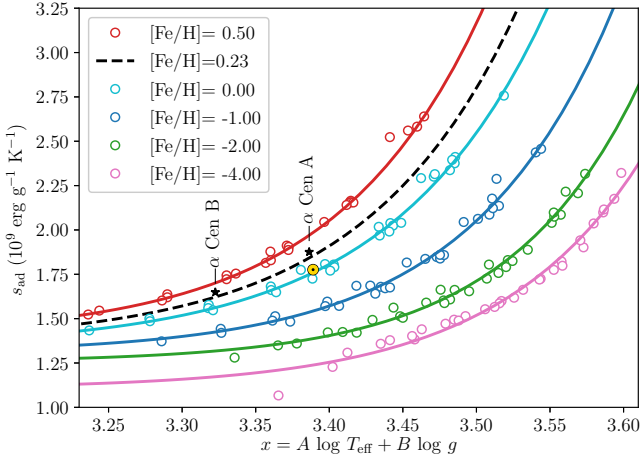}
\caption{Entropy calibration curves $s_{\rm ad}(x)$ at different metallicities (cf. Figure 3 of \citealt{Tanner_ea:2016}; open circles: 3D RHD simulations of \citealt{Magic_ea:2013b}; solid lines: analytic best fits of the results of the simulations, equations \ref{entropy} and \ref{xdef}); black stars: standard models of \acen{} A and B.}
\label{tanner}
\end{center}
\end{figure}

Furthermore, the results in Table \ref{entr_vs_std} imply that the accuracy of the determination of \sad{} and of the stellar radius are directly linked.
To clarify this point, Figure \ref{tanner} shows the analytic \sad{} vs. $x$ relations of \citet{Tanner_ea:2016} obtained as best fits of the 3D RHD simulations of \citet{Magic_ea:2013a,Magic_ea:2013b,Magic_ea:2015a,Magic_ea:2015b} and \citet{Tanner_ea:2013a,Tanner_ea:2013b,Tanner_ea:2014}, based on equations \eqref{entropy} and \eqref{xdef}.
The scatter of the open circles, representing the individual 3D RHD simulations, give a qualitative measure of the uncertainty of the fits.
The standard models of \acen{} A and B, and the \sad{} vs. $x$ relation interpolated at [Fe/H]$=0.23$, appropriate for the \acen{} system, are also plotted in Figure \ref{tanner}.
The standard models of \acen{} A and B have values of the adiabatic specific entropy that are consistent with the $s_{\rm ad}(x)$ curve interpolated at [Fe/H]$=0.23$ to the same level of accuracy of the fits, or better.

In conclusion, the main source of uncertainty in our entropy-calibrated models derives from that of the \sad{} vs. $x$ relations used.
An uncertainty of $\approx 1\%$ in \sad{} is intrinsic to the entropy calibration method in its current implementation, which is based on equations \eqref{entropy} and \eqref{xdef} with the coefficients determined by \citet{Tanner_ea:2016}.
Further improvement of the entropy calibration is conditional on the availability of a more accurate representation of the \sad{} vs. $x$ relation of the 3D RHD simulations.

\subsection{Evolutionary effects of the entropy calibration}

\begin{figure}
\begin{center}
\includegraphics[width=0.5\textwidth]{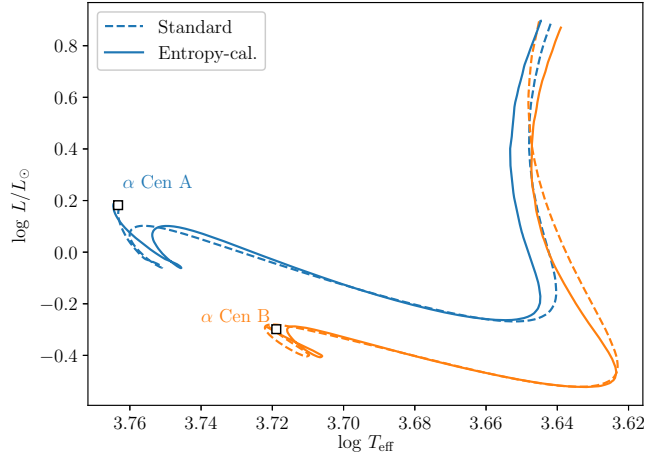}
\caption{Evolutionary tracks in the HR diagram of standard vs. entropy-calibrated models of \acen{} A and B. The data points are shown as empty squares, with the error bars being approximately the same size of the symbols.}
\label{evol}
\end{center}
\end{figure}

Figure \ref{evol} shows the evolutionary tracks in the HR diagram for the standard and entropy-calibrated models of \acen{} A and B.
The latter have Hayashi tracks displaced towards hotter $T_{\rm eff}$ by $\lesssim 70$ K with respect to the former. 
The shape and location of the main sequence portion of the tracks are also affected.

\begin{figure}
\begin{center}
\includegraphics[width=0.5\textwidth]{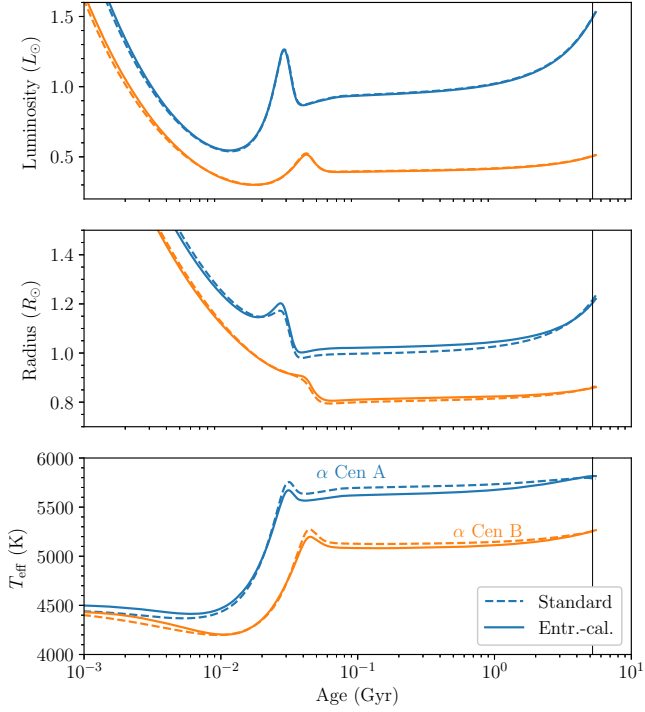}
\caption{Evolution of the standard vs. entropy-calibrated models of \acen{} A and B: luminosity (top panel); radius (middle), and effective temperature (bottom). The black vertical line marks the age of the system $t_A = t_B = 5.21$ Gyr.}
\label{teff_radius}
\end{center}
\end{figure}

The evolution of the luminosity, radius, and effective temperature is plotted in Figure \ref{teff_radius}.
Since the entropy-calibrated and the standard models share the same mass, age, and composition parameters, their differences are entirely the result of the different calibration of \cmlt{} in the two approaches.
The luminosity is very weakly sensitive to the entropy calibration of \cmlt{} (the difference in luminosity between the entropy-calibrated and standard models during the main sequence is less than $1\%$).
On the contrary, the radius is the most affected parameter ($\Delta R/R \lesssim 2$--$3\%$ for both stars). 
The effective temperature is also moderately affected.
As a consequence of $\Delta L/L \approx 0$, $\Delta T_{\rm eff}/T_{\rm eff} \approx -\frac{1}{2} \Delta R/R$.

\begin{figure}
\begin{center}
\includegraphics[width=0.5\textwidth]{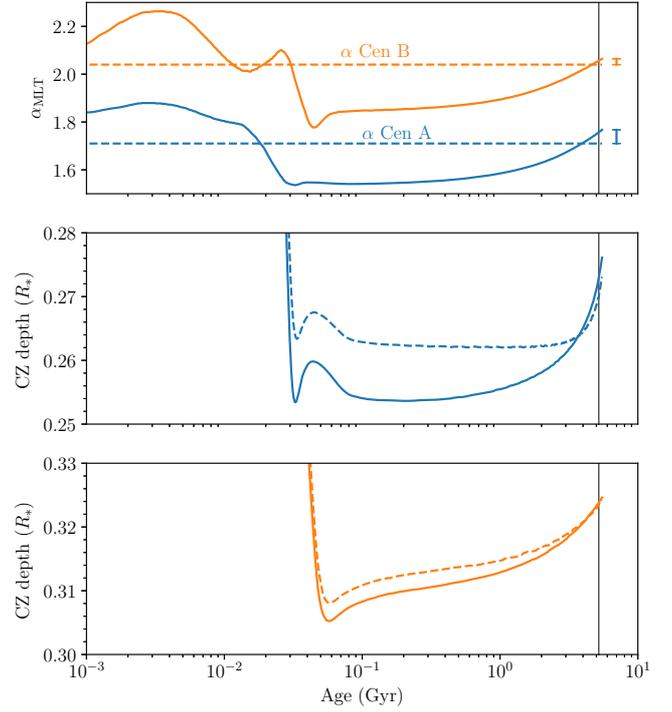}
\caption{Top panel: evolution of entropy-calibrated \cmlt{} (solid lines) compared with its constant value in the standard models (dashed lines). The difference in \cmlt{} at $t = t_A = t_B$ is $3.3\%$ and $1.2\%$ for \acen{} A and B, respectively. Middle and bottom panels: depth of the convection zone in stellar radii for for \acen{} A and B, respectively.}
\label{cmlt_cz}
\end{center}
\end{figure}

These differences are produced by the evolution of \cmlt{} in the entropy-calibrated runs, plotted in the top panel of Figure \ref{cmlt_cz}.
It should be noted that \cmlt{} at the age $t = t_A \approx t_B$ in the entropy-calibrated tracks does not exactly coincide with its constant value in  the standard tracks. 
The difference is more pronounced for \acen{} A than for \acen{} B ($\approx 3\%$ vs. $1\%$, respectively).
In contrast, for the entropy-calibrated solar models constructed in \citetalias{Spada_ea:2018} an exact agreement of \cmlt{} with the standard solar model at $t=t_\odot=4.57$ Gyr was imposed by introducing the additive constant term $s_{\rm offset}$ in equation \eqref{entropy}. 
The models of \acen{} A and B were constructed using the same value of $s_{\rm offset}$ as for the Sun, effectively anchoring the entropy calibration of \cmlt{} to the Sun. 

Apart from establishing the calibration of the stellar radius, the parameter \cmlt{} controls the depth of the convective envelope.
Indeed, the thickness of the convection zone differs significantly between the standard and entropy-calibrated models, as shown in the middle and bottom panels of Figure \ref{cmlt_cz}.
The difference between the standard and entropy-calibrated models is larger for \acen{} A.
Changes to the depth of the convection zone can affect the mixing of the outer layers, and have observable consequences in terms of the surface metallicity, or the abundance of light elements.

\begin{figure*}
\begin{center}
\includegraphics[width=\textwidth]{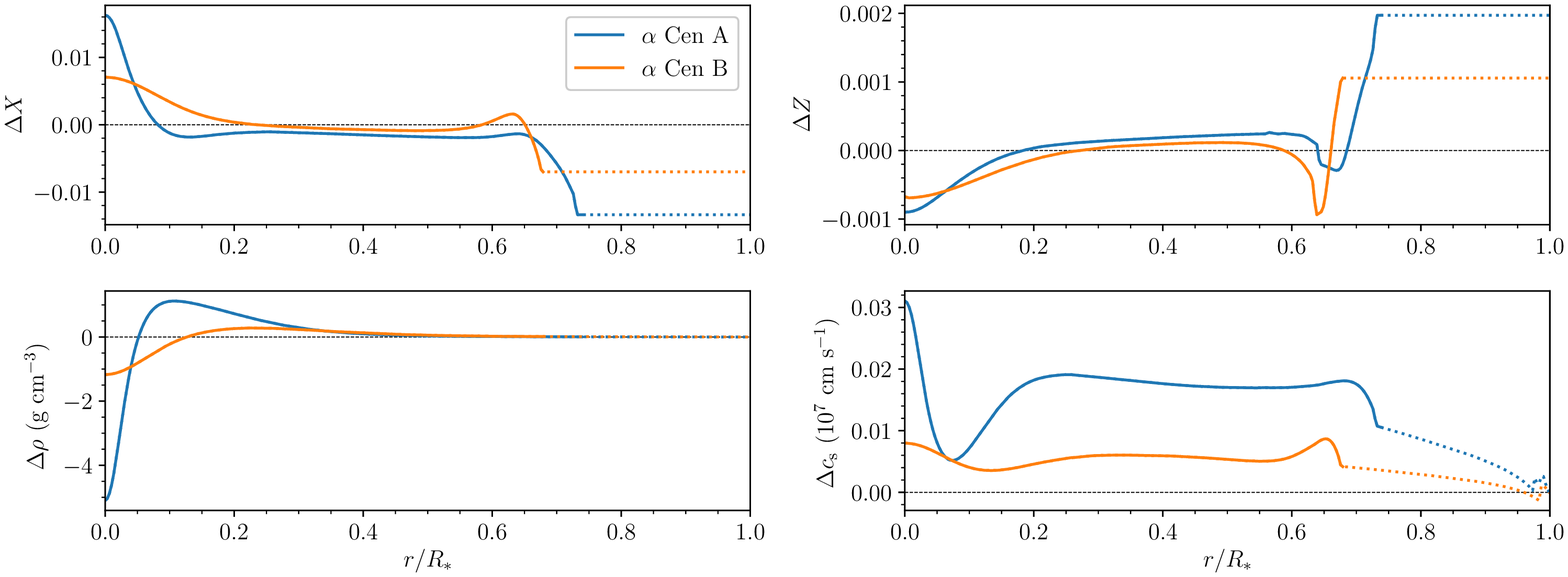}
\caption{Differences in the interior models of \acen{} A and B at age $t_A\approx t_B$. 
The panels show the radial profile of the difference between the entropy-calibrated and standard model in the specified variable. 
In all the plots, the dotted portion of the curves correspond to the convective envelope. 
Top left: hydrogen mass fraction; top right: bulk metallicity; bottom left: density; bottom right: sound speed.}
\label{int}
\end{center}
\end{figure*}

\subsection{Structural effects of the entropy calibration}

As a consequence of the cumulative effect of different mixing histories, the entropy-calibrated models have different interior composition profiles.
This is shown in the top panels of Figure \ref{int}. 
The differences in $X$ and $Z$ are most prominent in the convective envelope (which displays a flat profile characteristic of the efficient convective mixing), near the lower boundary of the convection zone, and close to the centre.
Note, in particular, the large difference (amounting to more than $10\%$) in the central hydrogen abundance of \acen{} A.  

In general, the differences in the density and sound speed profiles (bottom row plots of Figure \ref{int}) reflect both the different convection zone depth at the current age of the models, and the cumulative evolutionary effects, as the layers closest to the bottom of the convective envelope have undergone significantly different mixing in the entropy-calibrated models with respect to the standard ones.
In the entropy-calibrated model of \acen{} A, both the density and the sound speed differences clearly show the effect of the higher central hydrogen content.

%%%%%%%%%%%%%%%%%%%%%%%%%%%%%%%%%%%%%%%%%%%%
\section{Discussion}
\label{discussion}
%%%%%%%%%%%%%%%%%%%%%%%%%%%%%%%%%%%%%%%%%%%%

We have modelled the stars \acen{} A and B, comparing the standard approach, in which the MLT parameter is freely adjusted, with an entropy-based calibration of \cmlt{}, specified according to the results of 3D RHD simulations of convection.
The entropy calibration of \cmlt{}, recently proposed by \citet{Tanner_ea:2016}, is based on matching the adiabatic specific entropy obtained in the 1D stellar model with the corresponding value from a 3D RHD simulation of appropriate effective temperature, surface gravity, and metallicity.
This approach has been successfully used to construct a realistic solar model by \citet{Spada_ea:2018}.

It should be emphasised that the main goal of the entropy calibration approach is {\it not} to improve the estimate of the MLT parameter.
Indeed, the numerical value of \cmlt{} is well-known to be sensitive to the specific implementation of the MLT formalism, and is of little physical significance, since MLT does not provide a consistent description of convection (see, e.g., \citealt{Trampedach:2010}; see also \citealt{Tanner_ea:2016}).

In the entropy calibration method, the MLT formalism is adopted as a convenient procedure to provide more realistic boundary conditions for a 1D stellar model. 
Although the MLT does not reproduce the detailed structure and position of the SAL faithfully, the appropriate choice of \cmlt{} yields the correct entropy jump in the outer layers of a star.  
The specific entropy \sad{} in the adiabatic part of the convective envelope, in turn, determines the stellar radius.   
The accuracy of the radius thus depends on that of the entropy jump in the transition layer between the deep, optically thick layers, and the outer atmospheric layers (the SAL; see e.g. the discussions of \citealt{Straka_ea:2006, Kim_Chan:1997, Kim_Chan:1998}, and of \citealt{Tanner_ea:2014}). 
The value of \sad{} is not sensitive to the precise stratification and detailed structure of the SAL region and of the atmosphere: the density in both layers is low, and they comprise only a very small fraction of the envelope mass.  
Their combined geometrical extent and the details of their stratification contribute little to the total stellar radius.
As a result, the radii of individual 1D stellar models along an evolutionary track constructed with entropy-calibrated \cmlt{} are more accurate than those of standard models, even if the accuracy of the stratification of the SAL and of the outer layers is not improved by this approach.
In the entropy calibration method, the value of \cmlt{} is then used to map parametrically the dependence of \sad{} on the effective temperature, surface gravity, and metallicity derived from the RHD simulations in a form that can be readily incorporated into a 1D stellar evolution code.

The main ingredient of our entropy calibration of \cmlt{} is the calibration curve \sad{} vs. $x$, where $x = A\, \log T_{\rm eff} + B\, \log\, g$ (see equations \ref{entropy} and \ref{xdef} and Figure \ref{tanner}), which was derived by \citet{Tanner_ea:2016} from 3D RHD simulations of convection. 
The function $s_{\rm ad}(x)$ depends parametrically on the surface metallicity [Fe/H]. 
Remarkably, this function is insensitive to the details of the input physics and numerics of the 3D RHD simulations \citep[see][]{Tanner_ea:2016}, as well as those of the 1D stellar evolution code \citepalias[as was shown in][]{Spada_ea:2018}.
This property allows the construction of an accurate radius calibration for the 1D stellar models in terms of $s_{\rm ad}(x)$, even if the structure of the outer layers is calculated with the MLT formalism.
Such a calibration of \cmlt{} effectively removes one adjustable parameter in comparison with standard models.

Our entropy-calibrated models of \acen{} A and B reproduce the observed radii of both stars with an accuracy of $1\%$, or better, even if the freedom to adjust the value of \cmlt{} has been removed (all other parameters, such as masses, ages, and chemical composition, being the same). 
Together with the application to the solar model discussed in \citetalias{Spada_ea:2018}, the present work provides a solid test of the entropy calibration method.
It should be stressed that the accuracy on the stellar radius of the entropy-calibrated models is determined by that of the calibration curves $s_{\rm ad}(x)$.

While the entropy-calibrated model of \acen{} B reproduces its radius within the observational uncertainty, for \acen{} A the agreement is only to the $2.5\, \sigma$ level.
This result suggests that other effects, beyond the main contribution of \cmlt{}, can affect the two stars differently, and have a moderate ($\lesssim 1\%$) impact on their radii. 
The somewhat puzzling difference found in the best-fitting values of $Y_A$ and $Y_B$ (see Table \ref{entr_vs_std}) further corroborates this conjecture: it is possible that the different initial helium abundances effectively compensate for other uncertainties in the input physics.

For instance, the possibility that \acen{} A possesses a convectively unstable core has been considered by \citet{Joyce_Chaboyer:2018b}, as well as in earlier works. 
\citet{Bazot_ea:2016} have reported that including core overshooting leads to models of \acen{} A with a convective core.
Other microphysics choices, such as microscopic diffusion, or the rate of the $^{14}N(p,\gamma)^{15}O$ reaction, can also affect core convection. 
These authors have also found that models of \acen{} A with a convective core have, on average, slightly higher metallicity and helium abundance, as well as being younger and having a lower value of the MLT parameter.

To test the impact of core convection in \acen{} A on our conclusions, we have constructed entropy-calibrated models of this star that take core overshooting into account. 
Our treatment of core overshooting follows the standard YREC implementation \citep[e.g.,][]{Demarque_ea:2008,Spada_ea:2017}, in which the extent of the core overshoot region is proportional to an appropriate length scale, with the scaling factor given by the overshoot parameter $\alpha_{\rm OV}$. 
This length scale is equal either to the pressure scale height at the edge of the core, or, should this become unrealistically large because of the small size of the core, to the geometrical distance from the centre. 
It should be noted that in our models a short-lived episode of core convection is triggered around the ZAMS phase even without core overshooting. 
Values of the overshooting parameter $\alpha_{\rm OV}\lesssim 0.2$ extend the convective core size, but not its lifetime.
The convective core survives until the age of $\approx 5.2$ Gyr for $\alpha_{\rm OV} \approx 0.4$, which is probably too large to be realistic, see, e.g., Figure 1 of \citet{Spada_ea:2017}.
In any case, the entropy-calibrated model of \acen{} A with $\alpha_{\rm OV} = 0.4$ has a radius of $1.1967\, R_\odot$.
This is in slightly worse agreement with the observed value than the model without core overshooting. 
We conclude that core convection in \acen{} A, if present, does not explain the different level of accuracy of the radii of entropy-calibrated models.

Another possibility is that the lower accuracy for \acen{} A is due to the intrinsic limitation of the entropy calibration of \cmlt{}.
In particular, it is conceivable that a star of mass $\gtrsim 1.1\, M_\odot$, having a thinner convection zone than the Sun or \acen{} B, approaches the limit of applicability of the entropy calibration method.

Regardless of smaller effects, the capability of the entropy calibration method to produce stellar models with radii accurate within $1\%$ seems quite robust.
This level of accuracy is especially relevant for the characterisation of exoplanet-host stars in support of the exoplanet search missions ongoing or scheduled in the near future (e.g., TESS, PLATO).

%%%%%%%%%%%%%%%%%%%%%%%%%%%%%%%%%%%%%%%%%%%%
\section{Conclusions}
\label{conclusions}
%%%%%%%%%%%%%%%%%%%%%%%%%%%%%%%%%%%%%%%%%%%%

We have constructed models of \acen{} A and B implementing an entropy-based calibration of the MLT parameter based on 3D RHD simulations of convection, and compared them with standard models, where \cmlt{} is freely adjusted to reproduce the measured radius of each star.
Our main conclusions are as follows.

First, regarded as a test of our \cmlt{} calibration method, this comparison encouragingly shows that the entropy-calibrated models can reproduce the observed radii of \acen{} A and B with an accuracy of $1\%$ or better without ad-hoc tuning of \cmlt{}.

Second, the entropy-calibrated models can fit the radius of \acen{} A significantly less satisfactorily than that of \acen{} B. 
This result may be a hint of residual uncertainties in the modelling of stars close to the more massive end of the cool stars regime ($M \gtrsim 1.1\, M_\odot$), possibly related to core convection.
Alternatively, a lower accuracy for stars with thin outer convection zones (i.e., F vs. G or K spectral types) could be intrinsic to the entropy calibration itself, and, in turn, to the 3D RHD simulations from which it was derived. 
 
The results of this investigation and of \citetalias{Spada_ea:2018} validate the entropy calibration method, and open a promising way to integrate more realistic surface boundary condition, such as those provided by 3D RHD numerical simulations of convective envelopes, in 1D stellar evolution models.

\section*{Acknowledgements}
FS is supported by the German space agency (Deutsches Zentrum f\"ur Luft- und Raumfahrt) under PLATO Data Center grant 50OO1501.

\begin{figure*}
\begin{center}
\includegraphics[width=0.85\textwidth]{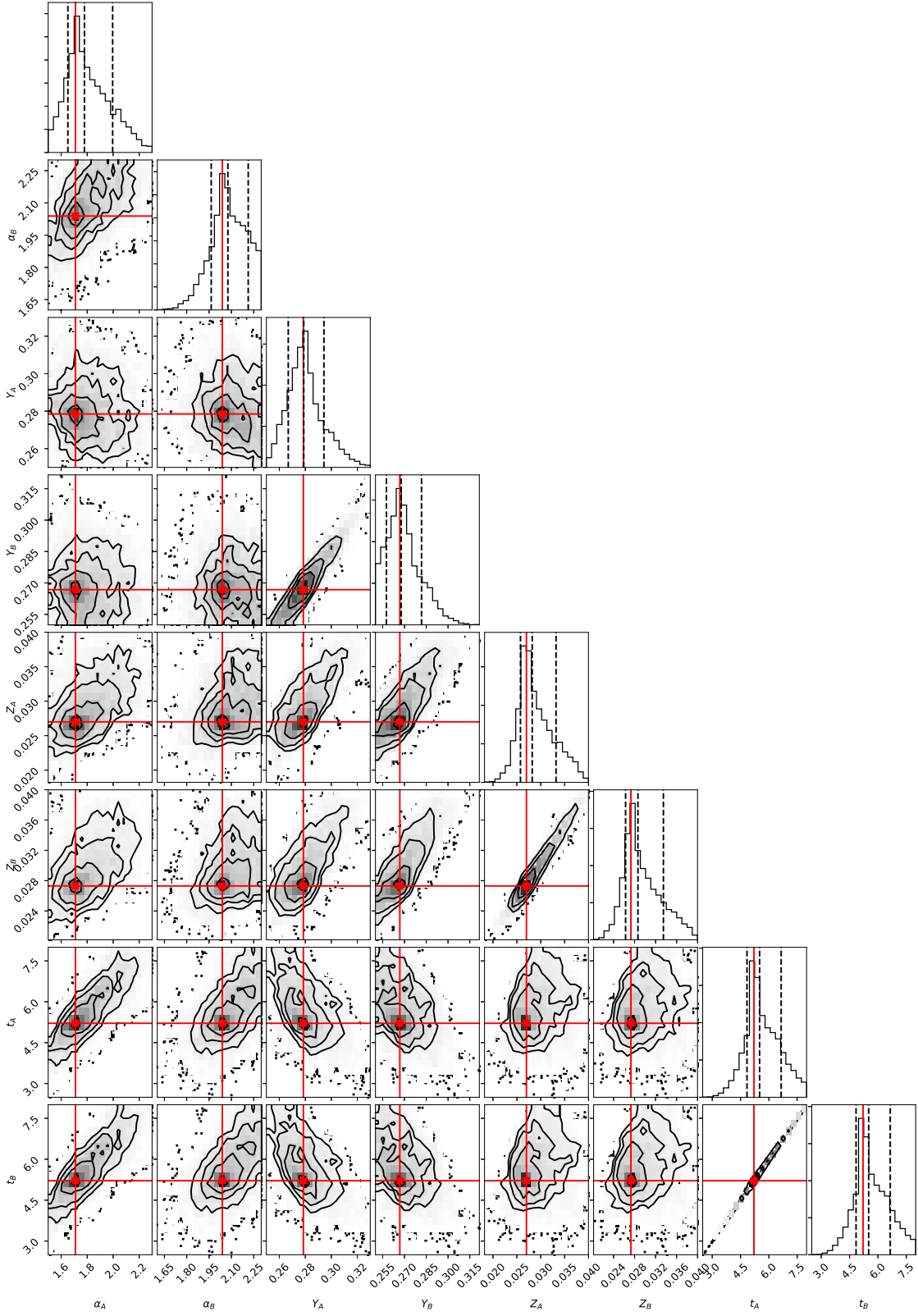}
\caption{``Corner plot", showing the posterior probability distributions and the correlations among the parameters of the best-fit with standard models of \acen{} A and B.}
\label{corner_plot}
\end{center}
\end{figure*}

\label{lastpage}
\end{document}